\documentclass[onecolumn,showpacs,preprintnumbers,amsmath,amssymb]{revtex4}

\usepackage{amssymb}
\usepackage{epsfig}
\usepackage{amsmath}

\begin{document}




\title{Alternative mechanism of avoiding the big rip or little rip for a scalar phantom field}


\author{Ping Xi}\email{xiping@shnu.edu.cn}
\author{Xiang-hua Zhai}\email{zhaixh@shnu.edu.cn}
\author{Xin-zhou Li}\email{kychz@shnu.edu.cn}
\affiliation{Shanghai United Center for Astrophysics(SUCA),
Shanghai Normal University, 100 Guilin Road, Shanghai 200234, P. R. China}

\begin{abstract}
Depending on the choice of its potential, the scalar phantom field $\phi$ (the equation of state parameter $w<-1$) leads to various catastrophic fates of the universe including big rip, little rip and other future singularity. For example, big rip results from the evolution of the phantom field with an exponential potential and little rip stems from a quadratic potential in general relativity (GR). By choosing the same potential as in GR, we suggest a new mechanism to avoid these unexpected fates (big and little rip) in the inverse-\textit{R} gravity. As a pedagogical illustration, we give an exact solution where phantom field leads to a power-law evolution of the scale factor in an exponential type potential. We also find the sufficient condition for a universe in which the equation of state parameter crosses $w=-1$ divide. The phantom field with different potentials, including quadratic, cubic, quantic, exponential and logarithmic potentials are studied via numerical calculation in the inverse-\textit{R} gravity with $R^{2}$ correction. The singularity is avoidable under all these potentials. Hence, we conclude that the avoidance of big or little rip is hardly dependent on special potential.
\end{abstract}



\maketitle
\section{Introduction}
It has been widely accepted that astrophysical observations might favor a dark energy component with "supernegative" equation of state (EOS) parameter $w=p/q<-1$, dubbed as phantom or super quintessence, where $p$ is the pressure and $\rho$ is the energy density \cite{Caldwell1,Hao1}. The phantom model with Born-Infeld type Lagrangian has been proposed \cite{Hao2} and its generalization to $O(n)$ symmetry has been done in Ref.\cite{Li1}. The phantom component leads to a catastropic big rip singularity that is characterized by divergences in the scale factor $a$, the Hubble parameter $H$ \cite{Caldwell2} and its time-derivative $\dot{H}$ \cite{Chimento} at the finite future $t=t_b$. It has been studied that various types of singularity might occur in phantom scenarios \cite{Nojiri1,Nojiri2}. Barrow and his collaborators \cite{Barrow1,Barrow2} have investigated general characterization of sudden cosmological singularities and the classical stability of cosmological solutions containing these singularities. Some mechanisms have been proposed to avoid these catastropic singularities. For example, it has been shown that these singularities in the future of the cosmic evolution might be avoided if we take suitable potential term \cite{Hao3,Hao4,Liu}, quantum effects \cite{Nojiri3} and modified gravity \cite{Elizalde}. If one considers an additional interaction between the phantom field and the background, the big rip singularity can be avoided in the late-time evolution of the universe \cite{Curbelo}. In brane-world models, a Gauss-Bonnet term is provided for the bulk, whereas phantom field is present on the brane. If the dark energy is modelled by a phantom-generalized chaplygin gas, the evolution of universe will avoid the big rip \cite{Hao5,Bouhmadi2,Zhai}. In the torsion cosmology, the scalar mode of torsion could be considered as a phantom field which will derive the universe in an oscillating fashion with an accelerating expansion at late-time \cite{Shie,Li2,Li3,Baekler,Ao}. However, $w<-1$ as the scale factor $a(t)$ extends into the future is necessary condition \cite{Nojiri1,Nojiri2}, but not sufficient one for a future singularity. Recently, the little rip scenario has been proposed in \cite{Frampton}, in which $w<-1$ but $w\rightarrow-1$ asymptotically, such that there is no future singularity. Such model will nevertheless lead to a dissolution of bound structures at some time in the future. Especially, the viscous fluid can produce a little rip cosmology as a purely viscosity effects \cite{Brevik}.

On the other hand, there is an interesting way to give rise to acceleration: a modification to the Einstein-Hilbert action involving new terms of inverse powers of the curvature scale (in the following text we call it inverse-\textit{R} gravity), of the form $\sqrt{-g}R^{-n}$ (with $n>0$) \cite{Carroll}. In general $f(R)$ gravity (even without scalar field), the divide $w=-1$ crossing behavior is shown in Refs. \cite{Nojiri4,Nojiri5,Bamba,Nojiri6}. Recently, Du et al. \cite{Du} discussed the behavior of quintessence field in the inverse-\textit{R} gravity and found that the evolution of $w$ with cosmic time can cross the divide $w=-1$ and $w$ will be less than $-1$. In this paper, we put forward a new mechanism to avoid the catastropic big or little rip by means of the investigation of phantom field in the inverse-\textit{R} gravity. It is worth noting that crossing $w=-1$ divide happens approximately in a few e-fold expansion when the inverse-\textit{R} term is still sufficiently large and the decrease of the energy density resulting from the crossing of $w$ will make the catastrophic fate avoided.

This paper is organized as follows. In Sec. 2, we give a brief review of inverse-\textit{R} gravity and its relation to the phantom field in Einstein's general relativity (GR). An exact solution is shown in Sec. 3. In Sec. 4, we investigate the mechanism for a single phantom field to avoid big or little rip by an analytical method. In Sec. 5, we present the numerical results to avoid big or little rip of phantom depending on different potentials in frame of inverse-\textit{R} gravity. Finally, we give a brief summary and a further discussion in Sec. 6.

\section{Phantom field in inverse-\textit{R} gravity}
There are two types of modification of Einstein-Hilbert action: one is ultraviolet modification which is important in the early universe (high curvature region) and the other is infrared modification which plays significant role in the late universe (low curvature region). The inverse-\textit{R} gravity is one of the current models with infrared modifications. The inverse-\textit{R} gravity seems unable to pass the solar system observations \cite{Chiba}, but this problem has been solved by adding a scalar curvature squared term $R^{2}$ to the inverse-\textit{R} gravity \cite{Nojiri7,Nojiri8}. In Ref. \cite{Nojiri6}, the following $f(R)$ theory was proposed:
\begin{equation}
f(R)=R-\frac{\alpha^{4}}{(R-\Lambda_1)^n}+(\frac{R-\Lambda_2}{\beta})^m
\end{equation}
where $n, m, \Lambda_1, \Lambda_2$ and $\beta$ are constants. This model leads to an acceptable cosmic speed-up and is consistent with some of the solar system observations. This general case returns to the inverse-\textit{R} gravity when the curvature is low and $\Lambda_1=0$.

In this paper, we consider the following action
\begin{equation}
S=\frac{m_{pl}^{2}}{2}\int d^{4}x\sqrt{-g}(R-\frac{\alpha^4}{R}+\frac{R^2}{\beta^2})+\int d^{4}x\sqrt{-g}(\mathcal{L}_m+\mathcal{L}_\phi)
\end{equation}
in the FRW universe with scale factor $a$. The metric is
\begin{equation}
ds^2=-dt^2+a^2d\Sigma^2
\end{equation}
where $d\Sigma^2$ is the metric of a 3-dimensional maximally symmetric space. $\mathcal{L}_m$ is the Lagrangian of matter and $\mathcal{L}_\phi$ is the Lagrangian of phantom field with the potential $V(\phi)$,
\begin{equation}
\mathcal{L}_\phi=\frac{1}{2}g^{\mu\nu}\partial_\mu\phi\partial_\nu\phi-V(\phi).
\end{equation}

The Friedmann equation is given by \cite{Nojiri6}
\begin{equation}
H^2+\frac{k}{a^2}-\frac{\mathcal{F}(H,\dot{H},\ddot{H})}{3m_{pl}^2}=\frac{\rho_m-\dot{\phi}^2/2+V(\phi)}{3m_{pl}^2}
\end{equation}
where $H\equiv\dot{a}/a$ is the Hubble parameter and the dot denotes the derivative with respect to $t$, and
\begin{eqnarray}
\mathcal{F}(H,\dot{H},\ddot{H})&=&\frac{\alpha^4m_{pl}^2}{12(\dot{H}+2H^2)^3}(2H\ddot{H}+15H^2\dot{H}+2\dot{H}^2+6H^4)\nonumber\\
&+&\frac{18m_{pl}^2}{\beta^2}(6H^2\dot{H}-\dot{H}^2+2H\ddot{H}).
\end{eqnarray}
The equation of motion for phantom field is
\begin{equation}
\ddot{\phi}+3H\dot{\phi}-\frac{dV}{d\phi}=0.
\end{equation}

In this system the phantom field together with the extra geometric term takes effect of dark energy. In the inverse-\textit{R} gravity, there is an additional $\mathcal{F}$-term in the modified Friedmann equation (5). To expound the observed evolving EOS of the effective dark energy, one can introduce the concept "equivalent dark energy" \cite{Du}. We derive the density of equivalent dark energy by comparing (5) with the standard Friedmann equation in GR. The latter can be written as
\begin{equation}
H^2+\frac{k}{a^2}=\frac{1}{3m_{pl}^2}(\rho_m+\rho_{de}).
\end{equation}
Comparing (8) with (5), one has
\begin{equation}
\rho_{de}=\rho_\phi+\mathcal{F},
\end{equation}
where $\rho_\phi=-\dot{\phi}^2/2+V(\phi)$. It is worth noting that $\mathcal{F}$-term makes a key difference from the standard model of GR. In the next section, we shall study analytically the phantom dynamics of universe in inverse-\textit{R} gravity.

\section{Power-law solution in a model with exponential type potential}
To find exact solutions is an important but difficult topic in such a highly non-linear system (7) and (10). The most interesting exact solution is a power-law one because it has the extra advantage that the equations for the generation of density perturbations can also be solved exactly in the cosmological inflation \cite{Stewart}. In a spatially flat universe, the Friedmann equation (5) reduces to
\begin{equation}
H^2-\frac{\mathcal{F}}{3m_{pl}^2}=\frac{-\dot{\phi}^2/2+V}{3m_{pl}^2}
\end{equation}
during the phantom dominated stage. Note that crossing $w=-1$ divide happens approximately in a few e-fold expansion when the inverse-\textit{R} term is still sufficiently large and the $R^2$-term is still sufficiently small, therefore, the $\beta$-term can be neglected in rhs of (6) at the late-time of evolution. We have
\begin{equation}
\mathcal{F}(H,\dot{H},\ddot{H})=\frac{\alpha^4m_{pl}^2}{12(\dot{H}+2H^2)^3}(2H\ddot{H}+15H^2\dot{H}+2\dot{H}^2+6H^4).
\end{equation}
In this case, the power-law solution arises when the potential is chosen to take the following form
\begin{equation}
V=V_0\exp{\Big[\frac{\phi}{m_{pl}}-\frac{\xi}{4}\ln{\Big|1+\gamma\exp{\Big(\frac{4\phi}{\xi m_{pl}}\Big)}\Big|}\Big]},
\end{equation}
where $\xi>4$, $\gamma$ and $V_0$ are constants, and
\begin{equation}
\gamma=\frac{3\alpha^4(12-3\xi-\sqrt{3\xi^2-12\xi})}{12(14\xi^3-64\xi^2-70\xi-17)\xi^2+4(24\xi^3-64\xi^2+37\xi-3)\xi\sqrt{3\xi^2-12\xi}}.
\end{equation}
The equations of motion (7) and (10) then have the particular solution
\begin{equation}
a=a_0t^j,
\end{equation}
\begin{equation}
\frac{\phi}{m_{pl}}=\xi\ln{t},
\end{equation}
where
\begin{equation}
j=\frac{\xi}{2}+\sqrt{\frac{\xi(\xi-4)}{12}}.
\end{equation}
Provided that $j>2$ when $\xi>4$, this solution satisfies the condition for accelerating expansion. If we take another value of $\gamma$ in the potential (12)
\begin{equation}
\gamma=\frac{3\alpha^4(12-3\xi+\sqrt{3\xi^2-4\xi})}{12(14\xi^3-64\xi^2-70\xi-17)\xi^2+4(24\xi^3-64\xi^2+37\xi-3)\xi\sqrt{3\xi^2-4\xi}},
\end{equation}
the system (7) and (10) have the particular solution (14) and (15), but
\begin{equation}
j=\frac{\xi}{2}-\sqrt{\frac{\xi(\xi-4)}{12}}.
\end{equation}
Provided that $1<j<2$ when $4<\xi<6$, this solution also satisfies the accelerating condition. It is not surprising that there is a power-law solution for the dynamical system of phantom because $\mathcal{F}$-term makes a key difference from the standard model of GR. The power-law solution makes a message known to us: the big rip can be avoided in the inverse-\textit{R} gravity.

\section{The avoiding mechanism}
Before studying the avoiding mechanism in this dynamical system, we discuss briefly the relation between density evolution and EOS. Since the dust matter obeys the continuity equation and the Bianchi identity keeps valid, equivalent dark energy itself also satisfies the continuity equation
\begin{equation}
\frac{d\rho_{de}}{dt}+3H(\rho_{de}+p_{eff})=0,
\end{equation}
where $p_{eff}$ denotes the effective pressure of the equivalent dark
energy. From (9) and (19), we can express EOS as
\begin{equation}
w=\frac{p_{eff}}{\rho_{de}}=-1-\frac{1}{3H(\rho_{\phi}+\mathcal{F})}(\dot{\rho_{\phi}}+\dot{\mathcal{F}})
\end{equation}
where
\begin{equation}
\dot{\mathcal{F}}=\frac{Hm_{pl}^2\alpha^4}{6(2H^2+\dot{H})^4}[\dddot{H}(2H^2+\dot{H})+3\ddot{H}(2H^2-7H\dot{H}-\ddot{H})+3\dot{H}(\dot{H}^2-16H^2\dot{H}-4H^4)]
\end{equation}
and $\dot{\rho_\phi}, \dot{\mathcal{F}}$ depend on the rate of expansion. If $\dot{\rho_\phi}+\dot{\mathcal{F}}<0$ ($\dot{\rho_\phi}+\dot{\mathcal{F}}>0$) the equivalent dark energy behaves as quintessence (phantom) in GR, i.e., $\rho_{de}$ decreases (increases) with expansion of universe.

Next we consider the approach of Ref.\cite{Nojiri1,Frampton} where the pressure is expressed as a function of the density in the form $p_{eff}=-\rho_{de}-f(\rho_{de})$. Taking a power law for $f(\rho_{de})$, i. e., $f(\rho_{de})=\frac{2A}{3}\rho_{de}^{\gamma}$ ($A>0$ and $\gamma>0$),we see that the fate of universe will depend on $\gamma$ : (i) There is the possibility of little rip but no finite future singularity for $\gamma \leq\frac{1}{2}$; (ii) There is a type I (big rip) singularity for $\frac{1}{2}<\gamma<1$; (iii) There exists a type III singularity for $\gamma>1$. Here, the singularities are classified according to Ref. [7]. We have
\begin{eqnarray}\label{11}
H= \left\{\begin{array}{ll}
 H_0 e^{A(t-t_0)},&\textrm{$\gamma=\frac{1}{2}$},\\
\frac{1}{\sqrt{3}}[(2\beta-1)A(t_s-t)]^{\frac{1}{1-2\gamma}},&\textrm{$\gamma>\frac{1}{2}$ and $\gamma\neq1$}
\end{array}\right.
\end{eqnarray}
where $\hbar=c=m_{pl}=1$ is taken for the sake of clarity. Substituting (22) into (21), we obtain
\begin{eqnarray}\label{11}
\dot{\mathcal{F}}= \left\{\begin{array}{ll}
 -\frac{6Ae^{-2A(t-t_0)}+21\sqrt{3}A^2e^{-3A(t-t_0)}+24A^3e^{-4A(t-t_0)}+3\sqrt{3}A^4e^{-5A(t-t_0)}}{[2+\sqrt{3}Ae^{-A(t-t_0)}]^4},&\textrm{$\gamma=\frac{1}{2}$},\\
-\frac{3M(t)N(t)}{2L(t)},&\textrm{$\gamma>\frac{1}{2}$ and $\gamma\neq1$}
\end{array}\right.
\end{eqnarray}
where
\begin{eqnarray}
L(t)&=&(\sqrt{3}+2(2\beta-1)(t_s-t)[A(2\beta-1)(t_s-t)]^{\frac{1}{1-2\gamma}})^4,\\\nonumber
M(t)&=&[A(2\beta-1)(t_s-t)]^{\frac{1}{2\gamma-1}},\\\nonumber
N(t)&=&2\sqrt{3}(1+2\beta)\beta+(2\beta-1)(t_s-t)[A(2\beta-1)(t_s-t)]^{\frac{1}{1-2\gamma}}(46\beta-16\beta^2-3)\\\nonumber
&+&4\sqrt{3}(2\beta-1)^2(t_s-t)^2[A(2\beta-1)(t_s-t)]^{\frac{2}{1-2\gamma}}\\\nonumber
&+&4(2\beta-1)^3(t_s-t)^3[A(2\beta-1)(t_s-t)]^{\frac{3}{1-2\gamma}}.
\end{eqnarray}
It is easy to prove $\dot{\mathcal{F}}<0$ ($\dot{\mathcal{F}}$ is a negative definite function) for $\gamma\leq\gamma_{crit}$, $\gamma_{crit}=(23+\sqrt{481})/16$, which brings a possibility to avoid the big and little rip, and some type III singularity for $\gamma\leq\gamma_{crit}$. In the inverse-\textit{R} gravity, the $\mathcal{F}$-term makes a key difference from the standard model of GR.

In the $w(t_0)\geq-1$ case, $w(t)$ will monotonically increase since $\dot{\mathcal{F}}$
 is a negative difinite function for $\gamma\leq\gamma_{crit}$. In the $w(t_0)<-1$ case, we need an analysis in detail. Combining (7) and (10), we find that
\begin{equation}
\dot{H}=\frac{3H\dot{\phi}^2+\dot{\mathcal{F}}}{6m_{pl}^2H},
\end{equation}
\begin{equation}
\ddot{H}=\frac{3\dot{H}\dot{\phi}^2+6H\dot{\phi}\ddot{\phi}-6m_{pl}^2\dot{H}^2+\ddot{\mathcal{F}}}{6m_{pl}^2H}.
\end{equation}
If $w(t_0)<-1$ from the current observation, we must demand that EOS parameter cross $w=-1$ divide for the avoidance of big or little rip in the future. The universal argument is that the crossing phenomenon happens at $t=t_c$ \textit{iff } $\dot{H}(t_c)=0$ and $\ddot{H}(t_c)\neq0$. In the phantom case of GR, we have $\dot{\mathcal{F}}\equiv0$ such that $\dot{H}(t_c)=0$ implies $\ddot{H}(t_c)=0$. Therefore, the crossing behavior never happens. In other words, the big or little rip is inescapable in the Einstein's GR. In inverse-\textit{R} gravity, $\dot{H}(t_c)=0$ and $\ddot{H}(t_c)\neq0$ \textit{iff} $\dot{\mathcal{F}}=-3H\dot{\phi}^2$ and $\ddot{\mathcal{F}}\neq-6H\dot{\phi}\ddot{\phi}$ at $t=t_c$. Thus, we have proved the following theorems.\\

Theorem. If $w(t_0)<-1$, the avoidance of big or little rip demands $\dot{\mathcal{F}}=-3H\dot{\phi}^2$ and $\ddot{\mathcal{F}}\neq-6H\dot{\phi}\ddot{\phi}$ at $t=t_c$ in the inverse-\textit{R} gravity where $t_c$ is the crossing time.\\

\section{Numerical calculation for the scalar phantom models in inverse-\textit{R} gravity with $R^2$ correction}

For a minimally coupled phantom field $\phi$ in GR, the necessary conditions have been given for the refrainment of big rip or little rip\cite{frampton2}. That is, when $V'/V\rightarrow\infty$ and $\int\sqrt{V}(V')^{-1}d\phi\rightarrow\infty$, $w$ approaches -1 sufficiently rapidly that big rip is refrained. Furthermore, the boundedness of the potential $V(\phi)$ indicates whether or not the model corresponds to a little rip expansion. Thus, we examine a specific model with $V(\phi)\rightarrow\infty$ as $\phi\rightarrow\infty$ in more detail.

In Sec. 4, we have investigated analytically the evolution of universe in the phantom dominated epoch in inverse-\textit{R} gravity. For a realistic universe, the pressureless matter is an essential component. However, it is difficult to study analytically the evolution of universe when we consider matter component in inverse-\textit{R} gravity with $R^2$ correction. The behaviors of phantom field with different potentials are investigated by numerical calculation in this section. The results show that the big or little rip can be avoided for these models. A spatially flat FRW universe has been assumed in all calculations.
\subsection{Exponential potential: An example of avoiding big rip}
The form of exponential potential is
\begin{equation}
V(\phi)=\mu_e\exp{(\frac{\phi}{m_{pl}})}
\end{equation}
where $\mu_e$ is a parameter with dimension of energy density. Let us introduce the four dimensionless parameters for the numerical calculations, $x\equiv\frac{H}{H_0}, y\equiv\frac{\phi}{m_{pl}}, b\equiv\frac{\alpha^4}{H_0^4}$ and $c_0\equiv\frac{H_0^2}{\beta^2}$. Thus, (7) and (10) can be rewritten as
\begin{equation}
x^2=\Omega_{m,0}e^{-3s}-\frac{1}{6}x^2y'^2-u(y)+b(x)+c(x),
\end{equation}
\begin{equation}
x^2y''+xx'y'+3x^2y'-v(y)=0,
\end{equation}
where
\begin{equation}
b(x)=\frac{b(4x^2x'^2+2x^3x''+15x^3x'+6x^4)}{36(xx'+2x^2)^3},
\end{equation}
\begin{equation}
c(x)=18c_0(6x^3x'+x^2x'^2+2x^3x''),
\end{equation}
\begin{equation}
u(y)=\frac{\mu_ee^y}{3H_0^2m_{pl}^2},
\end{equation}
\begin{equation}
v(y)=\frac{\mu_ee^y}{H_0^2m_{pl}^2}.
\end{equation}
Here the prime denotes the derivative with respect to the so called e-folds $s\equiv\ln{a}$ and $\Omega_{m,0}\equiv\rho_{m,0}/(3H_0^2m_{pl}^2)$, $H_0$ is the current Hubble parameter and $\rho_{m,0}$ is the current density of pressureless matter. Generally speaking, $c_0$ is a little tiny number since the inflation scale $\beta$ is much higher than $H_0$. If we fix the inflation scale as a GUT energy scale of $10^{14}$ GeV, $c_0\approx10^{-112}$.

The deceleration parameter can be expressed as
\begin{equation}
q=-1-\frac{x'}{x}
\end{equation}
and we introduce the dimensionless density of equivalent dark energy $\Omega_{de}$ for convenience
\begin{equation}
\Omega_{de}=\frac{\rho_{de}}{3H_0^2m_{pl}^2}=-\frac{1}{6}x^2y'^2+u(y)+b(x)+c(x).
\end{equation}
Using numerical calculation, we plot $w, \Omega_{de}$ and $q$ as functions of e-folds $s$ for the scalar phantom in inverse-\textit{R} gravity in Fig. 1. In Fig. 2, we show the evolutions of $w$ for the scalar phantom model with an exponential potential in GR and in inverse-\textit{R} gravity. Obviously, the EOS of equivalent dark energy changes from less than $-1$ to larger than $-1$ such that the big rip singularity is avoidable in the inverse-\textit{R} gravity whereas $w$ is always less than $-1$ and asymptotically tends to a constant which leads to the big rip singularity in GR. It is clear that $R^{-1}$-term plays an essential role for the avoidance of big rip. The result of numerical calculation is consistent with that of exact power law solution Eqs. (14) and (15) for exponential type potential.
\begin{figure}
\epsfig{file=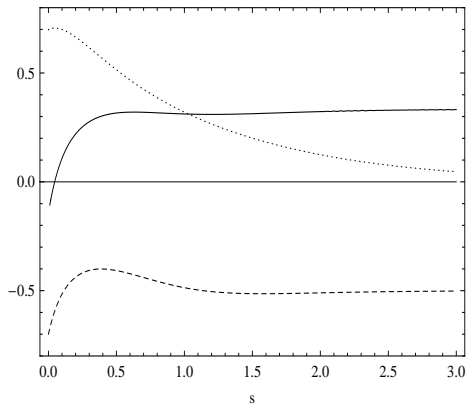,height=2.1in,width=2.5in}\caption{The evolution of $1+w, \Omega_{de}$ and $q$ with the e-folds $s$ for an exponential potential, where the real, dot, dash lines represent $1+w, \Omega_{de}$ and $q$, respectively.}
\end{figure}

\begin{figure}
\epsfig{file=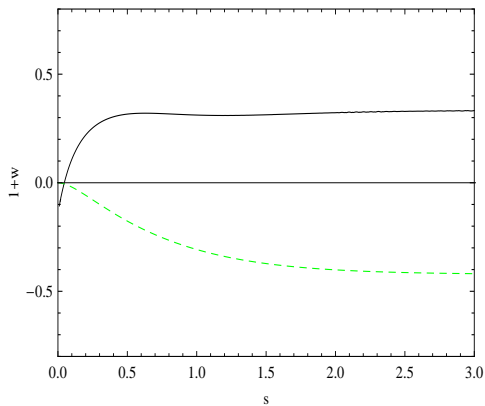,height=2.1in,width=2.5in}\caption{The behaviors of $1+w$ for the scalar phantom model with an exponential potential in GR (the dash line)and in inverse-\textit{R} gravity (the real line).}
\end{figure}

\subsection{Quadratic potential: An example of avoiding little rip}
The quadratic potential is widely investigated since any potential around its minimum (if it has a minimum) can be treated as the form of quadratic. We take the potential as
\begin{equation}
V(\phi)=\mu_2^2\phi^2
\end{equation}
where $\mu_2$ is a parameter with dimension of mass. In this case, (7) and (10) are still reduced to (28) and (29), but
\begin{equation}
u(y)=\frac{\mu_2^2}{3H_0^2}y^2,
\end{equation}
\begin{equation}
v(y)=\frac{2\mu_2^2}{H_0^2}y.
\end{equation}

In GR, phantom model with  quadratic potential may lead to the little rip in the future of cosmic evolution. However, we see that the EOS of equivalent dark energy changes from less than $-1$ to larger than $-1$ in Fig. 3. It means that the little rip is avoidable for phantom field with a quadratic potential in the inverse-\textit{R} gravity.
\begin{figure}
\epsfig{file=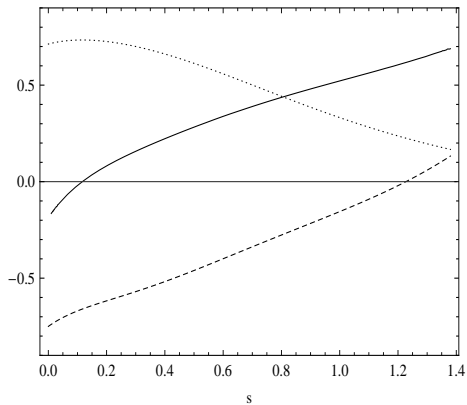,height=2.1in,width=2.5in}\caption{The evolution of $1+w, \Omega_{de}$ and $q$ with the e-folds $s$ for the quadratic potential, where the real, dot, dash lines represent $1+w, \Omega_{de}$ and $q$, respectively. The initial values are taken as $x_0=1, y_0=1, x_0'=-1.03, y_0'=0.01$.}
\end{figure}

\subsection{Other potentials}
In this subsection, we study other potentials for phantom field including cubic, quantic and logarithmic potentials. In GR, the scalar phantom field with quadratic, cubic and quantic potentials might lead to little rip whereas the big rip may occur for exponential and logarithmic potentials\cite{frampton2}. In the numerical calculation, we only need to revise the expressions of $u(y)$ and $v(y)$ in (28), (29) and (35) for different potentials, which are shown in Table 1.
\begin{table}[thbp]
\caption{The expressions of $u(y)$ and $v(y)$ for various potentials}
\begin{tabular}{c c c}
\hline
$V(\phi)$ \qquad &$u(y)$ \qquad&$v(y)$ \qquad\\
\hline
$\mu_2^2\phi^2$& $\frac{\mu_2^2y^2}{3H_0^2}$ & $\frac{2\mu_2^2y}{H_0^2}$\\
$\mu_3\phi^3$& $\frac{\mu_3m_{pl}y^3}{3H_0^2}$ & $\frac{3\mu_3m_{pl}y^2}{H_0^2}$\\
$\mu_4\phi^4$& $\frac{\mu_4m_{pl}^2y^4}{3H_0^2}$ & $\frac{4\mu_4m_{pl}^2y^3}{H_0^2}$\\
$\mu_e\exp{(\frac{\phi}{m_{pl}})}$& $\frac{\mu_ee^y}{3H_0^2m_{pl}^2}$ & $\frac{\mu_ee^y}{H_0^2m_{pl}^2}$\\
$\mu_l\ln{(\frac{\phi}{m_{pl}})}$& $\frac{\mu_l\ln{y}}{3H_0^2m_{pl}^2}$ & $\frac{\mu_l}{H_0^2m_{pl}^2y}$\\
\hline
\end{tabular}
\end{table}

In Figs. 4-6, we show the numerical results for the cubic, quantic and logarithmic potentials, respectively. We find that the results are analogous to the case of exponential or quadratic potential, i.e., the big or little rip is avoidable in the inverse-\textit{R} gravity.
\begin{figure}
\epsfig{file=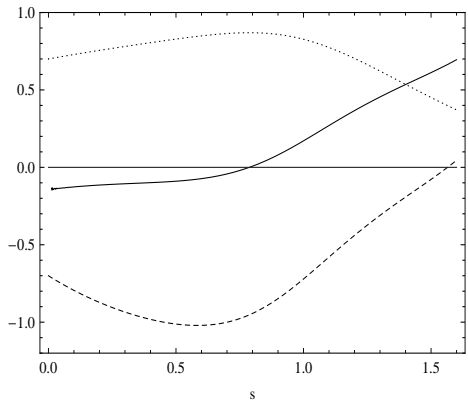,height=2.1in,width=2.5in}\caption{The evolution of $1+w, \Omega_{de}$ and $q$ with the e-folds $s$ for the cubic potential, where the real, dot, dash lines represent $1+w, \Omega_{de}$ and $q$, respectively.}
\end{figure}
\begin{figure}
\epsfig{file=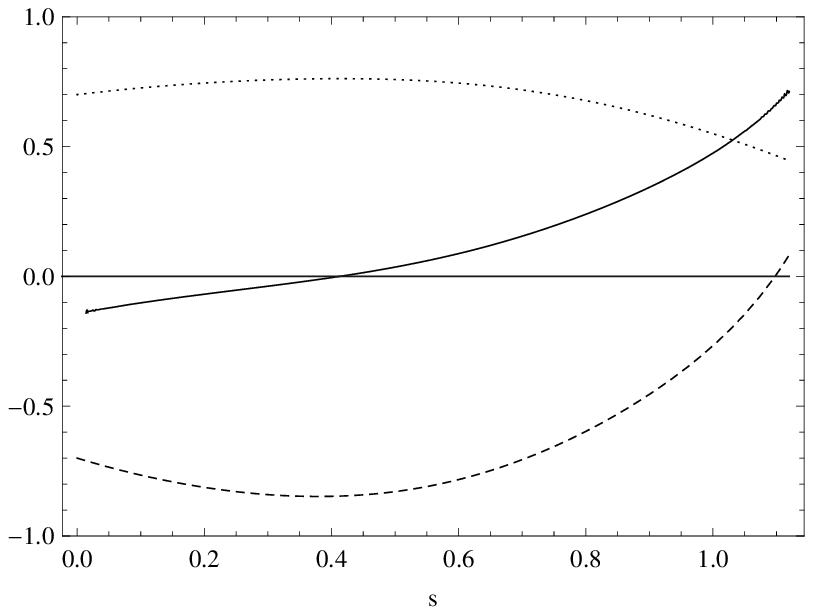,height=2.1in,width=2.5in}\caption{The evolution of $1+w, \Omega_{de}$ and $q$ with the e-folds $s$ for the quantic potential, where the real, dot, dash lines represent $1+w, \Omega_{de}$ and $q$, respectively.}
\end{figure}
\begin{figure}
\epsfig{file=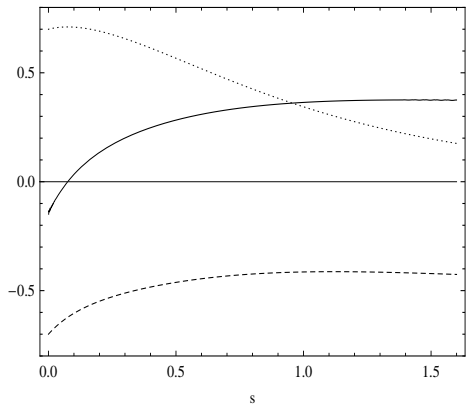,height=2.1in,width=2.5in}\caption{The evolution of $1+w, \Omega_{de}$ and $q$ with the e-folds $s$ for the logarithmic potential, where the real, dot, dash lines represent $1+w, \Omega_{de}$ and $q$, respectively.}
\end{figure}

\section{Conclusion}
In GR, the phantom field with an exponential or a quadratic potential leads to big or little rip that is characterized by the destruction of all bound structures. We show the possibility of avoiding the big or little rip naturally in the inverse-\textit{R} gravity. For a special exponential type potential, we find an exact power-law solution that makes a message known to us: the big rip can be avoided in the inverse-\textit{R} gravity. We also find the sufficient condition for a universe in which the equation of state parameter crosses $w=-1$ divide in the inverse-\textit{R} gravity. In GR, the scalar phantom field with quadratic, cubic and quantic potentials might lead to little rip whereas the big rip may occur for exponential and logarithmic potentials. And then we investigate these potentials for phantom field $\phi$ in the inverse-\textit{R} gravity with $R^2$ correction. The big or little rip is avoidable under all these potentials. Therefore, we conclude that the avoidance of big or little rip is a robust property of inverse-\textit{R} gravity with $R^2$ corretion, not rigidly adhered to some special potential.

It is worth noting that crossing $w=-1$ divide happens approximately in a few e-fold expansion when the inverse-\textit{R} term is still sufficiently large and the decrease of the energy density resulting from the crossing of $w$ will make the catastrophic fate avoided.

Finally, we give a brief discussion. It is not a shortcoming that we discuss only the background evolution in this paper but not consider the cosmological perturbation theory. In fact, Nojiri and Odintsov\cite{Nojiri6} have pointed out that modified gravity equations of motion are higher-derivative differential equations. The cosmological perturbations in modified gravity approaches to those in GR, since the corresponding equations are reduced to second-order differential equations.

\noindent {\bf Acknowledgments}\\
This work is supported by National Education Foundation of China
under grant No. 200931271104, National Science Foundation of China under Grant. Nos. 11075106 and 10671128, Innovation Program of Shanghai Municipal Education Commission (11zz123) and Key Project of Chinese Ministry of Education (No. 211059).

\end{document}